\pdfoutput=1

\documentclass[11pt,twoside,a4paper,cmspaper,final,collab]{cms-tdr}
\def\svnVersion{316224}\def\cmsCernNo{2013-037}\def\cmsCernDate{\today}\def\cmsMessage{Published in Physics Letters B as \href{http://dx.doi.org/10.1016/j.physletb.2015.09.057}{\doi{10.1016/j.physletb.2015.09.057.}}}

\begin{document}\cmsNoteHeader{HIN-13-007}

\hyphenation{had-ron-i-za-tion}
\hyphenation{cal-or-i-me-ter}
\hyphenation{de-vices}
\RCS$Revision: 316177 $
\RCS$HeadURL: svn+ssh://svn.cern.ch/reps/tdr2/papers/HIN-13-007/trunk/HIN-13-007.tex $
\RCS$Id: HIN-13-007.tex 316177 2015-12-28 13:46:55Z alfloren $
\newlength\cmsFigWidth
\ifthenelse{\boolean{cms@external}}{\setlength\cmsFigWidth{0.98\columnwidth}}{\setlength\cmsFigWidth{0.65\textwidth}}
\ifthenelse{\boolean{cms@external}}{\providecommand{\cmsLeft}{top\xspace}}{\providecommand{\cmsLeft}{left\xspace}}
\ifthenelse{\boolean{cms@external}}{\providecommand{\cmsRight}{bottom\xspace}}{\providecommand{\cmsRight}{right\xspace}}
\providecommand {\rootsNN}  {\ensuremath{\sqrt{s_{_\mathrm{NN}}}}\xspace}
\providecommand{\pPb}{\ensuremath{\mathrm{pPb}}\xspace}
\newcommand{\etalab}{\eta_\text{lab}}
\hyphenation{a-na-ly-sis}
\cmsNoteHeader{HIN-13-007}
\title{Study of W boson production in pPb collisions at $\rootsNN=5.02$\TeV}

\date{\today}

\abstract{The first study of W boson production in pPb collisions is presented, for bosons decaying to a muon or electron, and a neutrino. The measurements are based on a data sample corresponding to an integrated luminosity of 34.6\nbinv at a nucleon-nucleon centre-of-mass energy of $\rootsNN=5.02$\TeV, collected by the CMS experiment. The W boson differential cross sections, lepton charge asymmetry, and forward-backward asymmetries are measured for leptons of transverse momentum exceeding 25\GeVc, and as a function of the lepton pseudorapidity in the $\abs{\etalab}< 2.4$ range. Deviations from the expectations based on currently available parton distribution functions are observed, showing the need for including W boson data in
 nuclear parton distribution global fits.
}

\hypersetup{%
pdfauthor={CMS Collaboration},%
pdftitle={Study of W boson production in pPb collisions at sqrt(s[NN]) = 5.02 TeV},%
pdfsubject={CMS},
pdfkeywords={CMS, W boson, pPb collisions}
}

\maketitle

\section{Introduction}

Electroweak boson production in proton-nucleus and nucleus-nucleus collisions at the CERN LHC offers a unique opportunity to probe nuclear parton distribution functions (nPDFs)~\cite{Kartvelishvili:1995fr,Vogt:2000hp,Zhang:2002yz,Paukkunen:2010qg}. Leptonic decays of electroweak bosons are of particular interest since leptons do not interact strongly with
the medium produced in these collisions~\cite{ConesadelValle:2007sw,ConesadelValle:2009vp}.
As compared to those in a proton, the nPDFs are expected to be depleted (\emph{shadowing}) for partons carrying small momentum fractions $x\lesssim10^{-2}$, and enhanced (\emph{anti-shadowing}) in the $5\times 10^{-2} \lesssim x \lesssim10^{-1}$ range~\cite{Armesto:2006ph}.
However, because of the lack of available data, parton densities are less precisely known for nuclei than for nucleons.  As a consequence, precise calculations describing hard processes in high-energy heavy ion collisions are limited by uncertainties in the nPDFs.
 For \PW\ boson production, the dominant processes at LHC energies are
$\cPqu\cPaqd \to \PWp$ and $\cPqd \cPaqu \to \PWm$, principally reflecting interactions that take place between valence quarks and sea antiquarks. According to Ref.~\cite{Paukkunen:2010qg}, PDF nuclear modifications could affect the yield of \PW\ bosons in pPb collisions at the LHC by as much as 15\% in certain kinematic regions.
Therefore, precise measurements of \PW\ boson production in
heavy ion collisions might lead to an improved determination of the nPDFs. Moreover, asymmetries in the individual yields of \PWp\ and \PWm\ should permit the flavour decomposition of \cPqu\ and \cPqd\ quark distributions in nuclei.

The ATLAS~\cite{Aad:2010aa,Aad:2012ew} and
CMS~\cite{Chatrchyan:2011ua,Chatrchyan:2014csa} collaborations reported the observations of \cPZ\
bosons in heavy ion interactions, at a centre-of-mass energy of 2.76\TeV per nucleon pair. These data showed that the Z boson yields per nucleon--nucleon (NN) collision are essentially
unmodified by the medium produced in the collisions.
Although \PW\ bosons decaying to a lepton and a neutrino are more difficult to detect, their rate is
about ten times larger than that of \cPZ\ bosons decaying to leptonic final states.
The production of \PW\ bosons in PbPb collisions was reported by CMS~\cite{Chatrchyan:2012nt} and ATLAS~\cite{Aad:2014bha}, using data corresponding to an integrated luminosity of 7.3~$\mu$b$^{-1}$ and 150~$\mu$b$^{-1}$, collected in 2010 and 2011, respectively.
The W boson yield per NN collision was shown to be compatible with the one measured in pp collisions, when taking into account isospin effects arising from the mixture of protons and neutrons in the colliding nuclei.
However, the presence of 10--20\% nPDF effects on \cPZ\ and \PW\ boson production could not be excluded due to the relatively large experimental and theoretical uncertainties of these results.

The 2013 pPb LHC run provides the best currently available data sample to look for initial-state effects (such as PDF modifications) using electroweak bosons. The NN-equivalent luminosity is of the same order of magnitude as for the 2011 PbPb run, and the production cross sections are approximately a factor of two greater owing to the increased energy, 5.02\TeV per nucleon pair. Furthermore, the
asymmetry of the pPb collision system allows for the measurement of other observables such as forward-backward pseudorapidity asymmetries.
This Letter reports a study of \PW\ boson production in a sample of pPb collisions corresponding to an integrated luminosity of $(34.6 \pm
1.2)\nbinv$~\cite{CMS-PAS-LUM-13-002}, collected by the CMS
experiment.

\section{Experimental methods}

The direction of the proton beam was initially opposite to the positive direction of the CMS longitudinal axis~\cite{Chatrchyan:2008zzk}, and was reversed after 60\% of the data were taken.
The beam energies were 4\TeV for protons and 1.58\TeV per nucleon for lead nuclei, resulting in a centre-of-mass energy per
nucleon pair of $\rootsNN = 5.02$\TeV.  As a result of the energy difference of the
colliding beams, the NN centre-of-mass frame in pPb collisions was not at rest with respect to the laboratory
frame. Massless particles emitted at pseudorapidity  $\eta$
in the NN centre-of-mass frame are detected at
$\etalab = \eta - 0.465$ (first proton beam orientation) and
 $\etalab = \eta + 0.465$  (second proton beam orientation) in the CMS coordinate
system, as defined in Ref.~\cite{Chatrchyan:2008zzk}. The results presented hereafter are expressed in the usual convention of the proton-going side defining the positive pseudorapidity.
It coincides with the CMS convention in the second period of data taking, the first one being reversed before summing yields from the two beam configurations.

A detailed description of the CMS detector can be found elsewhere~\cite{Chatrchyan:2008zzk}. Its central feature is a superconducting solenoid of 6\unit{m} internal diameter, providing a magnetic field of 3.8\unit{T}. Within the field volume are the silicon  pixel-and-strip tracker, a lead tungstate crystal electromagnetic calorimeter (ECAL), and a brass and scintillator hadron calorimeter (HCAL), each composed of a barrel and two endcap sections.
The silicon tracker consists of 66\unit{M} pixel and 10\unit{M} strip sensor elements, and measures charged-particle trajectories in the pseudorapidity range $\abs{\etalab} < 2.5$.
Outside of the solenoid, muons are detected in the $\abs{\etalab} < 2.4$ range, with gas-ionization detector planes based on three technologies: drift tubes, cathode strip chambers, and resistive-plate chambers.
Electrons are identified in the ECAL, which is made of 75\,848 lead tungstate crystals and covers $\abs{\etalab}< 1.48 $ in the barrel and $1.48 < \abs{\etalab} < 3.00$ in the two endcap regions.
The CMS apparatus also has extensive forward calorimetry, including two steel/quartz-fiber Cherenkov hadron forward (HF) calorimeters, which cover the $2.9 < \abs{\etalab} < 5.2$ range.
For online event selection, CMS uses a two-level trigger system.

Selection criteria similar to the ones developed in Ref.~\cite{Chatrchyan:2013nka} are applied to the pPb sample to
remove events with electromagnetic, beam-gas, or multiple collisions (pileup). 
The W boson yields are corrected for the induced $(4.0\pm0.5)\%$ signal loss. 

The primary signature of a $\PW$ boson is a high transverse momentum (\pt) lepton. The current analysis is restricted to leptons of \pt greater than 25\GeVc.
The muon analysis is based on a sample triggered by requiring a single
muon with \pt above 12\GeVc, while the electron analysis uses an ECAL-triggered sample with a transverse energy
threshold of 15\GeV.
Leptons are reconstructed with the same algorithms as in proton-proton collisions~\cite{MuonReco,Chatrchyan:2013dga}, and standard selection criteria are applied, as in Refs.~\cite{Chatrchyan:2012nt,CMS:2013hoa}.
A special electron charge determination,  as described in Ref.~\cite{Chatrchyan:2012xt}, is used in
order to reduce the electron charge misidentification to a sub-percent level.
Events are reconstructed using particle-flow (PF) techniques~\cite{CMS-PAS-PFT-09-001,CMS-PAS-PFT-10-001}, which reconstruct and classify individual particles with an optimised combination
of all subdetector information.

Two criteria are used to remove specific background sources. First, events with two oppositely charged leptons, with the second lepton \pt greater than 15\,(10)\GeVc for muons (electrons) are removed, since they correspond to well-identified processes like Drell--Yan, \cPZ\ boson or high-\pt quarkonium production.
Second, the leptons are required to be isolated, in order to reduce the contamination coming from jet fragmentation.
The energies of all PF candidates are summed within a cone centred around the lepton, with the exception of the lepton itself. The lepton is considered isolated if the total transverse energy in the cone is small compared to its transverse momentum.
For muons, a cone of radius $\Delta R = \sqrt{\smash[b]{(\Delta\eta)^2+(\Delta\phi)^2}} = 0.3$ is used, where $\Delta\eta$ and $\Delta\phi$ are the pseudorapidity and azimuthal distances to the lepton. The candidate is rejected if the in-cone transverse energy is greater than 10\% of the muon~\pt. For electrons, a cone of $\Delta R = 0.4$ is used, and only particles with \pt greater than 1\GeVc are summed, to reduce the underlying-event enhanced contribution. The electron candidate is rejected if the resulting transverse
energy is greater than 11.5\%\,(9.5\%) of the electron \pt, for the ECAL barrel (endcaps).

\begin{figure*}[hbtp]\centering
\includegraphics[width=0.45\textwidth]{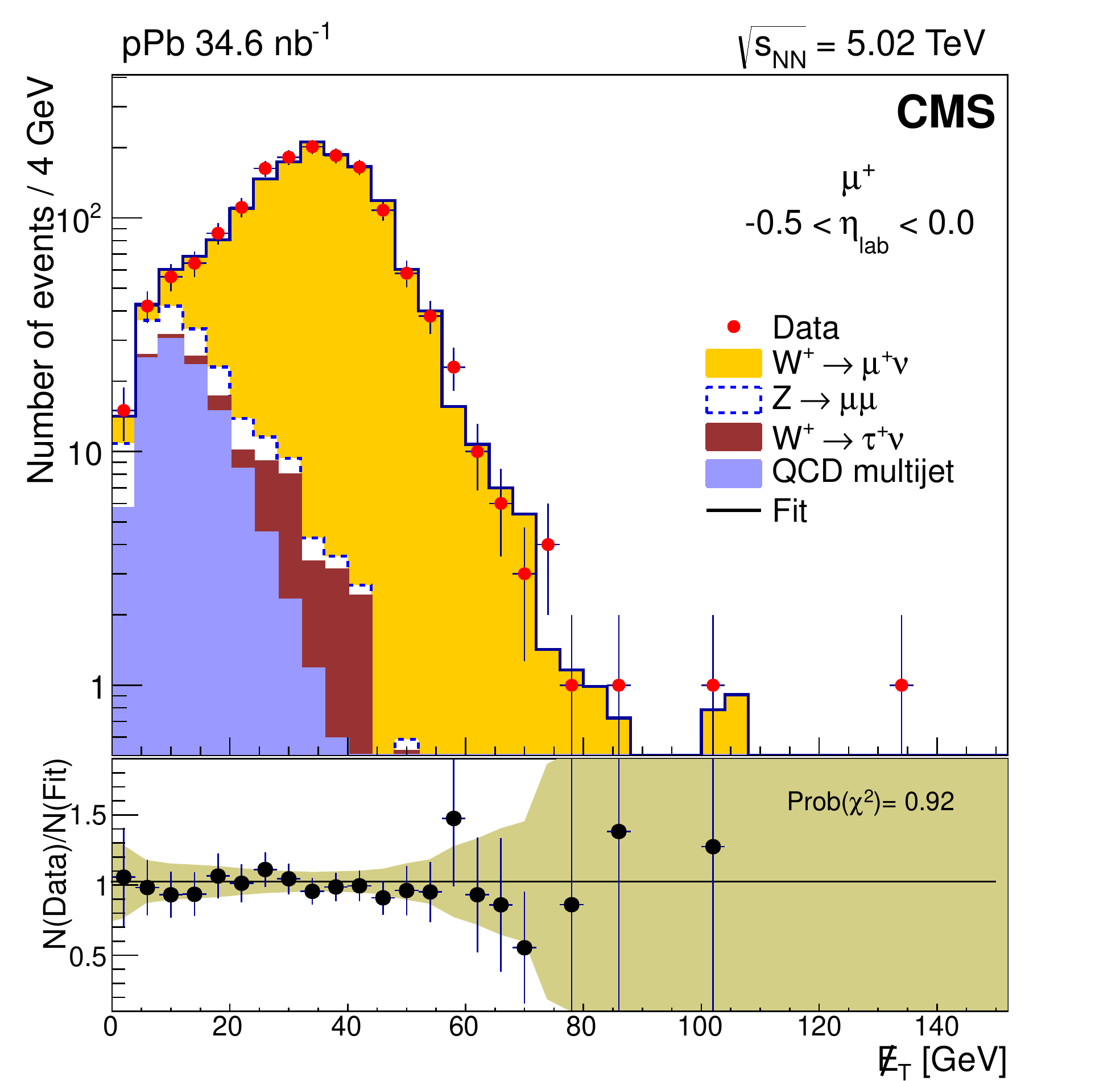}
\includegraphics[width=0.45\textwidth]{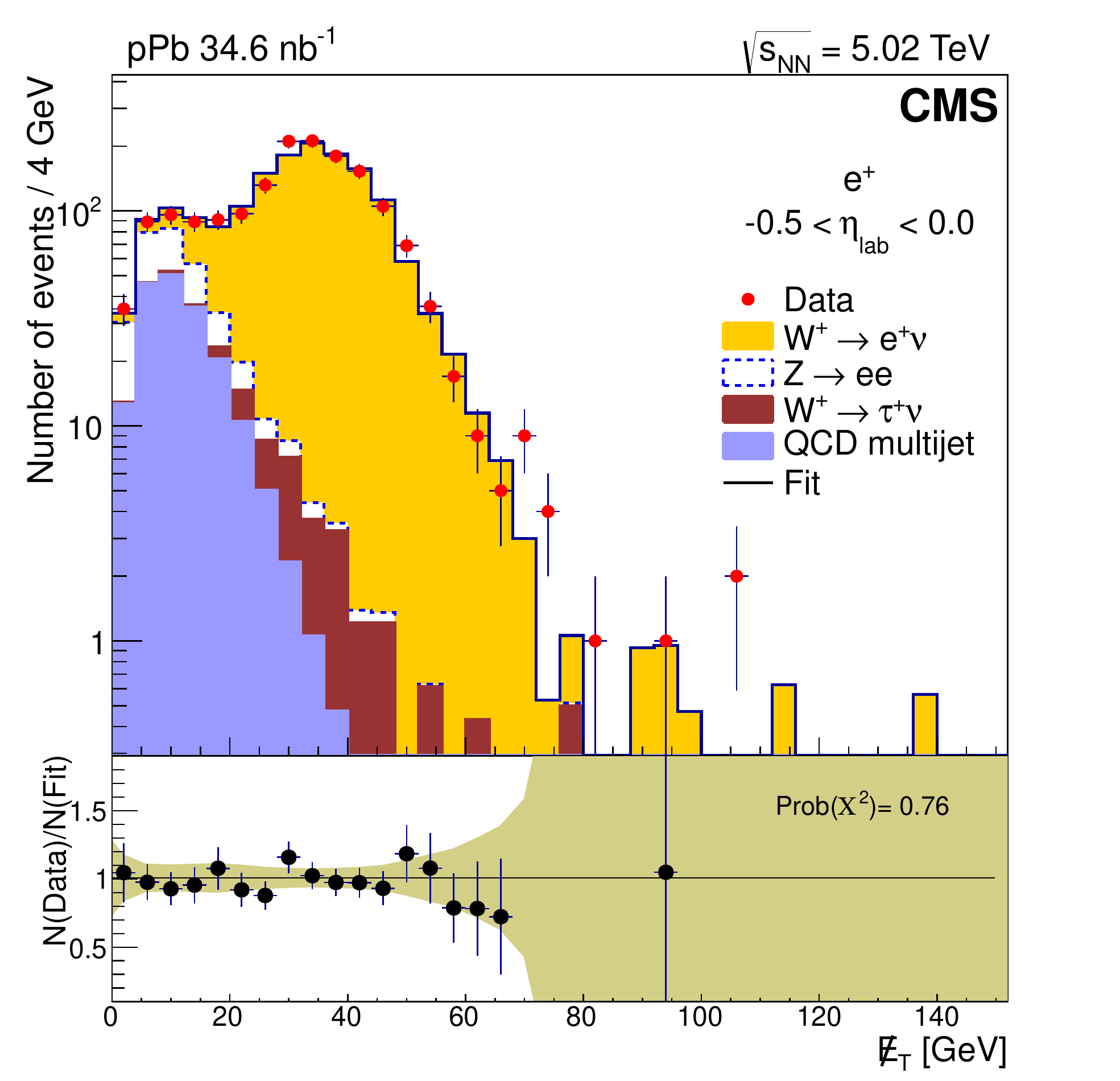}
\includegraphics[width=0.45\textwidth]{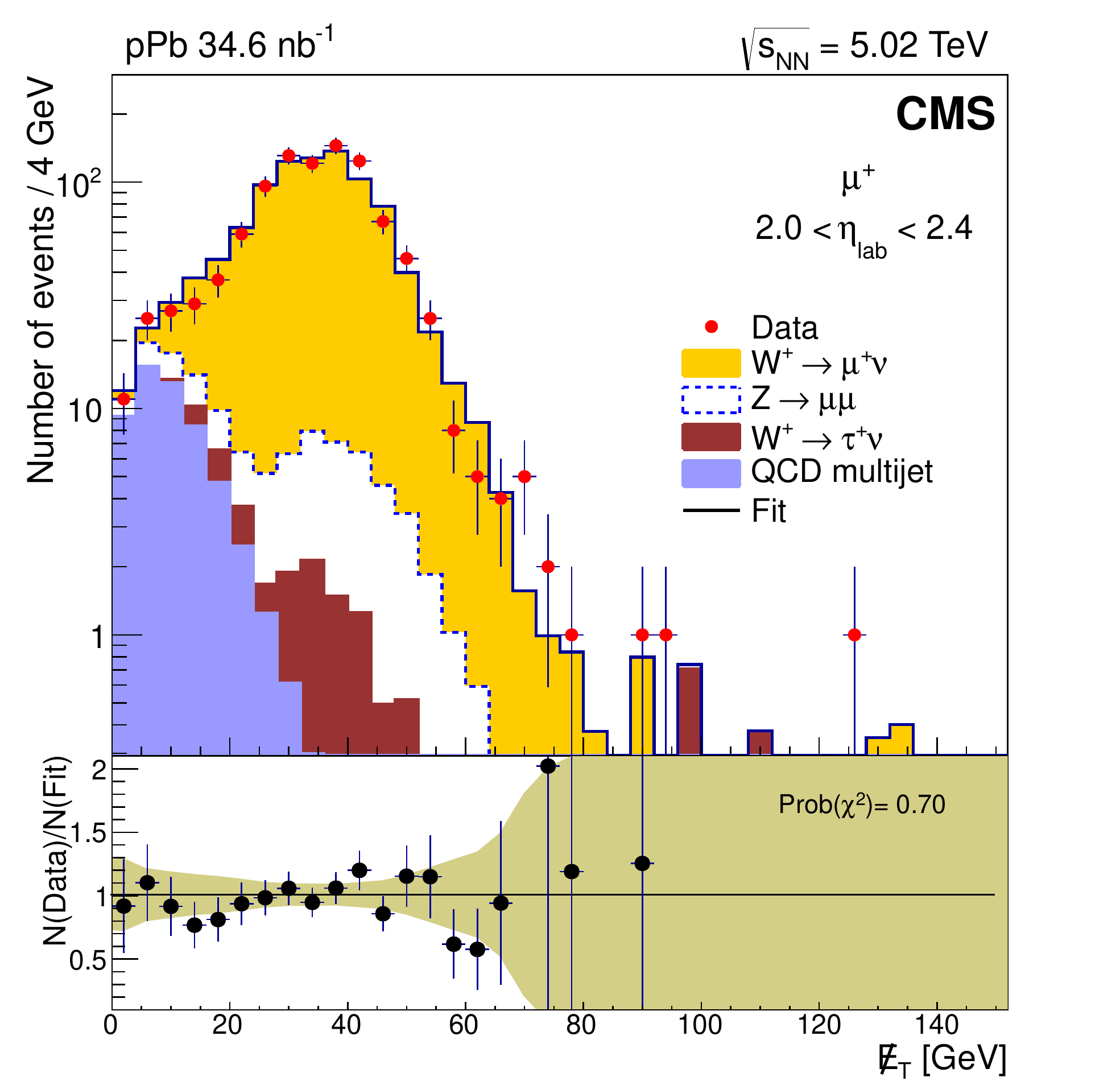}
\includegraphics[width=0.45\textwidth]{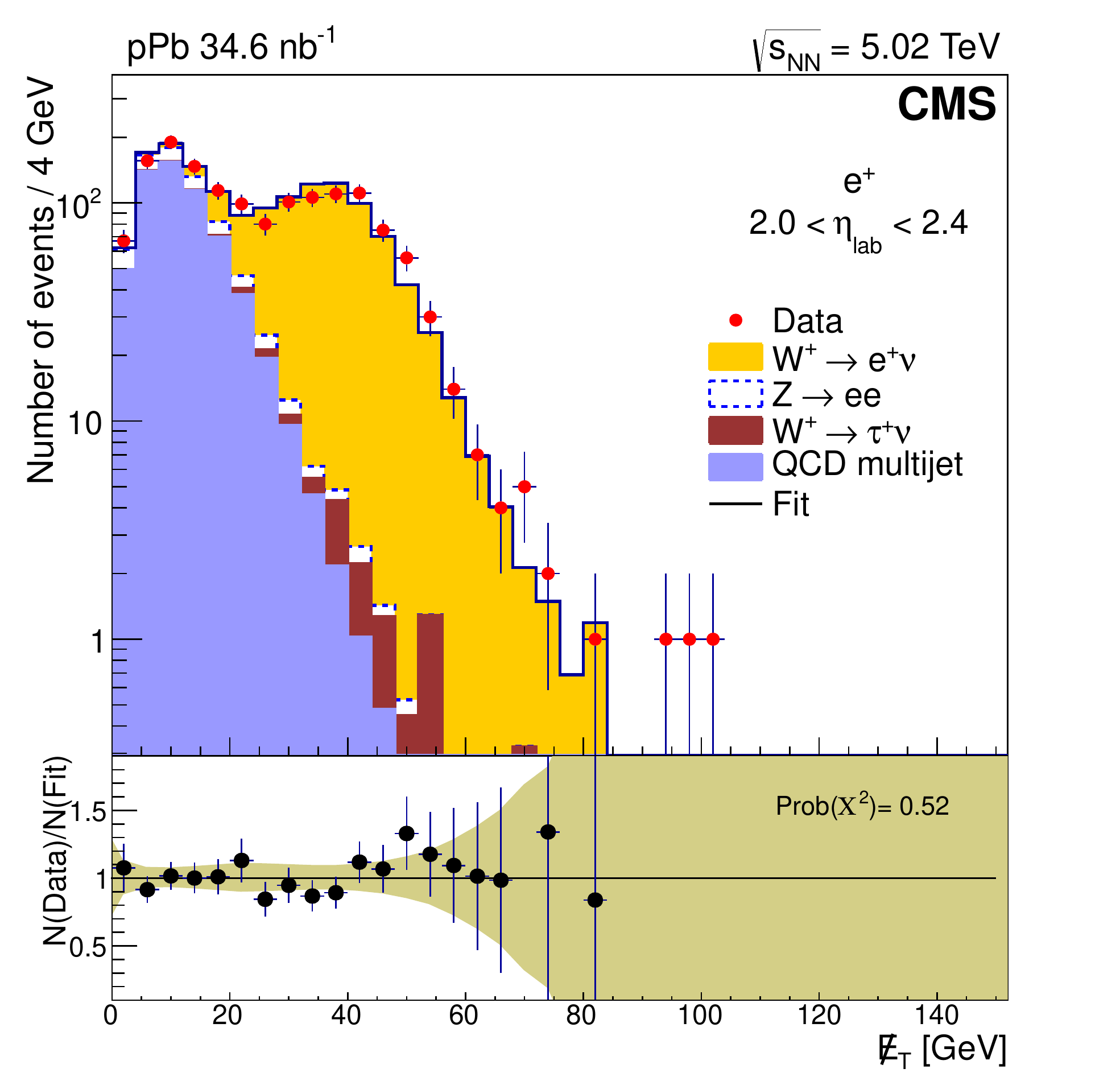}
 \caption{Missing transverse energy distribution for $\PWp \to \Pgmp \nu$ (left) and $\PWp \to \Pep \nu$ (right) events within the $-0.5 < \etalab < 0.0$ (top) and $2.0 < \etalab < 2.4$ (bottom) ranges. Binned fits to the data (red points) are performed with four contributions, stacked from bottom to top: multijet (QCD, blue), $\PWp \to \tau^+ \nu$ (brown), $\cPZ \to \ell \ell$ (white) and $\PWp \to \ell^+ \nu$ (yellow). The $\etalab$ regions are defined such that the proton is moving towards positive $\etalab$ values. Error bars represent statistical uncertainties. The lower panels display the data divided by the result of the fit, with the band representing the statistical uncertainties on the sum of the fit components, for each \ETslash bin. }  \label{fig:MET}\end{figure*}

An important characteristic of events containing a $\PW \to \ell \nu$ decay is the missing transverse energy (\ETslash) associated with the undetected neutrino. It is computed as the magnitude of the vectorial sum of transverse momenta of all the PF candidates in the event.
The analysis is performed using ten lepton pseudorapidity bins, each 0.5 wide except for the most forward and backward regions ($2<\abs{\etalab}<2.4$). After having applied the lepton selection criteria, examples of the resulting \ETslash distributions are shown in Fig.~\ref{fig:MET} for $\Pgmp$ and $\Pep$, in the most central ($-0.5 < \etalab < 0.0$) and furthest forward
($2.0 < \etalab < 2.4$)
ranges. The distributions for other bins and for the negative leptons are similar.

To extract the number of events with a lepton coming from a $\PW$ boson,
binned fits of these distributions are performed, including the signal and main background contributions, in each $\etalab$ bin. The \ETslash shapes assumed for the electroweak processes,
namely the $\PW^{\pm} \to \ell^{\pm} \nu$ signal as well as background from $\PW^{\pm} \to \tau^{\pm} \nu$ and $\cPZ \to \ell^+\ell^-$, are determined by the simulations described hereafter, taking into account the acceptance and efficiency. Their relative normalization is given by the unmodified
theoretical cross sections (as computed in Ref.~\cite{CMS:2011aa}). A maximal 20\% variation of the $\PW/\cPZ$ normalization ratio is taken into account, due to potentially different nuclear modifications of the \cPZ\ and \PW\ bosons, and resulting in a 1--3\% systematic uncertainty in the extracted \PW\ yields. The noticeable difference between the \ETslash distributions for the $\cPZ \to \Pep \Pem$ and $\cPZ \to \Pgmp \Pgmm$ processes
in the forward region (bottom plots of Fig.~\ref{fig:MET}) results from the greater ECAL coverage allowing missed electrons with $2.4<\abs{\etalab}<3.0$ to be accounted for in the \ETslash calculation.
The shape of the QCD multijet background is modelled by the functional form $f(\ETslash)=(\ETslash+\ETslash^0)^\alpha \exp (\beta \sqrt{\ETslash+\ETslash^0})$. It is shown to reproduce the \ETslash shape of data events containing non-isolated leptons, with the $\ETslash^0$, $\alpha$, and $\beta$ parameters, which are observed to depend mildly and linearly on the cone/lepton transverse energy ratio. These fitted parameters are then extrapolated to the isolated lepton signal regime and the resulting function is used as the QCD background shape.
The multijet background contribution is larger in the electron channel because the misidentified lepton rate is higher, particularly due to a contribution from photon-jet events. Contributions from other sources, such as $\cPqt  \cPaqt$ production and high-\pt quarkonia, were found to be negligible. 

A small charge misidentification correction (less than 0.2\%) is applied to the electron yields; this correction is negligible for muons. All fits are of good quality, as illustrated by the bottom panels of Fig.~\ref{fig:MET} that show the ratio of the data to the fit outcome. The observed numbers of leptons coming from \PW\ boson decays over the entire pseudorapidity range are: $11660 \pm 111$~$\Pgmp$, $9459 \pm 99$~$\Pgmm$, $9892 \pm 116$~$\Pep$, and $7872 \pm 101$~$\Pem$, where the uncertainty is statistical, determined by the fit procedure.

In order to correct for inefficiencies in the lepton trigger, reconstruction, and selection, the electroweak processes $\PW
\to \ell \nu$ have been simulated using the \PYTHIA 6.424 generator~\cite{Sjostrand:2006za} with a mixture of pp and
pn interactions corresponding to pPb collisions.
 The detector response to each \PYTHIA signal event is simulated with \GEANTfour~\cite{Agostinelli:2002hh} and then
embedded in a minimum bias pPb background event. These background events are produced with the \textsc{hijing} event
generator~\cite{Gyulassy:1994ew} and passed through \GEANTfour as well. Each simulation is done twice, once for each proton beam direction, and includes a boost to reproduce the 0.465 rapidity shift. The embedding is done at the level of
detector hits, and the signal and background events share the same generated vertex location.
The embedded event is then processed through the trigger emulation and the full event reconstruction chain.
The resulting reconstructed events are then reweighted to match the distributions observed in data of the event vertex and activity (as measured in the HF calorimeters).
The obtained efficiencies vary with $\etalab$ (with higher efficiencies at mid-rapidity), from 59\% to 89\% for muons, and from 51\% to 84\% for electrons.

The various components of the single-lepton efficiency are also directly computed from pPb data, using $\cPZ \to
\ell \ell$ samples, and techniques described in Ref.~\cite{CMS:2011aa}. These efficiencies are then compared to the
corresponding efficiencies computed from simulations. In the case of trigger and reconstruction efficiencies, they are found to be consistent. The
isolation criterion rejects more leptons in data, because the local activity of the underlying event is greater than in the
simulation. To account for such discrepancies, the efficiency from $\PW \to \ell \nu$ simulation is
multiplied by correction factors, which are determined as the ratio of the single-lepton efficiencies measured in $\cPZ \to
\ell \ell$ data to those estimated in simulations. The so-called ``tag-and-probe'' method used for this estimation is described in Ref.~\cite{Khachatryan:2010xn}.
These correction factors are computed in bins of $\etalab$ and for positively and negatively charged muons separately. In the electron case, the low statistical precision motivates a correction factor estimated for electrons and positrons combined.

The total systematic uncertainty in the lepton yields is estimated by adding the different contributions in quadrature. The $\etalab$-dependent sources of systematic uncertainty arise from the method used for the estimation of multijet
background (0.1--2.0\% for muons, 0.5--3.8\% for electrons), the normalization of the electroweak background (1--3\% for muons and electrons), the efficiency correction factors (2.2--7.5\% for muons, 2.6--7.4\% for electrons), and the energy scale of electrons (0.1--2.0\%). The uncertainty in the momentum scale of muons is found to be negligible.
The integrated luminosity measurement uncertainty (3.5\%~\cite{CMS-PAS-LUM-13-002}) affects
only the W boson production cross sections and cancels in the asymmetry measurements, as does the additional global uncertainty arising from the efficiency of the filter rejecting pileup events (0.5\% for both channels). Though the common electron/positron correction factors cancel, a residual systematic uncertainty of 3\% is assigned to the charge asymmetry, based on simulation studies and $\etalab$-integrated efficiencies determined from $\cPZ \to \Pep \Pem$ data. No other systematic uncertainty cancellations are assumed for the asymmetry results.

\section{Results}

\begin{figure}[hbtp]\centering
 \includegraphics[width=0.45\textwidth]{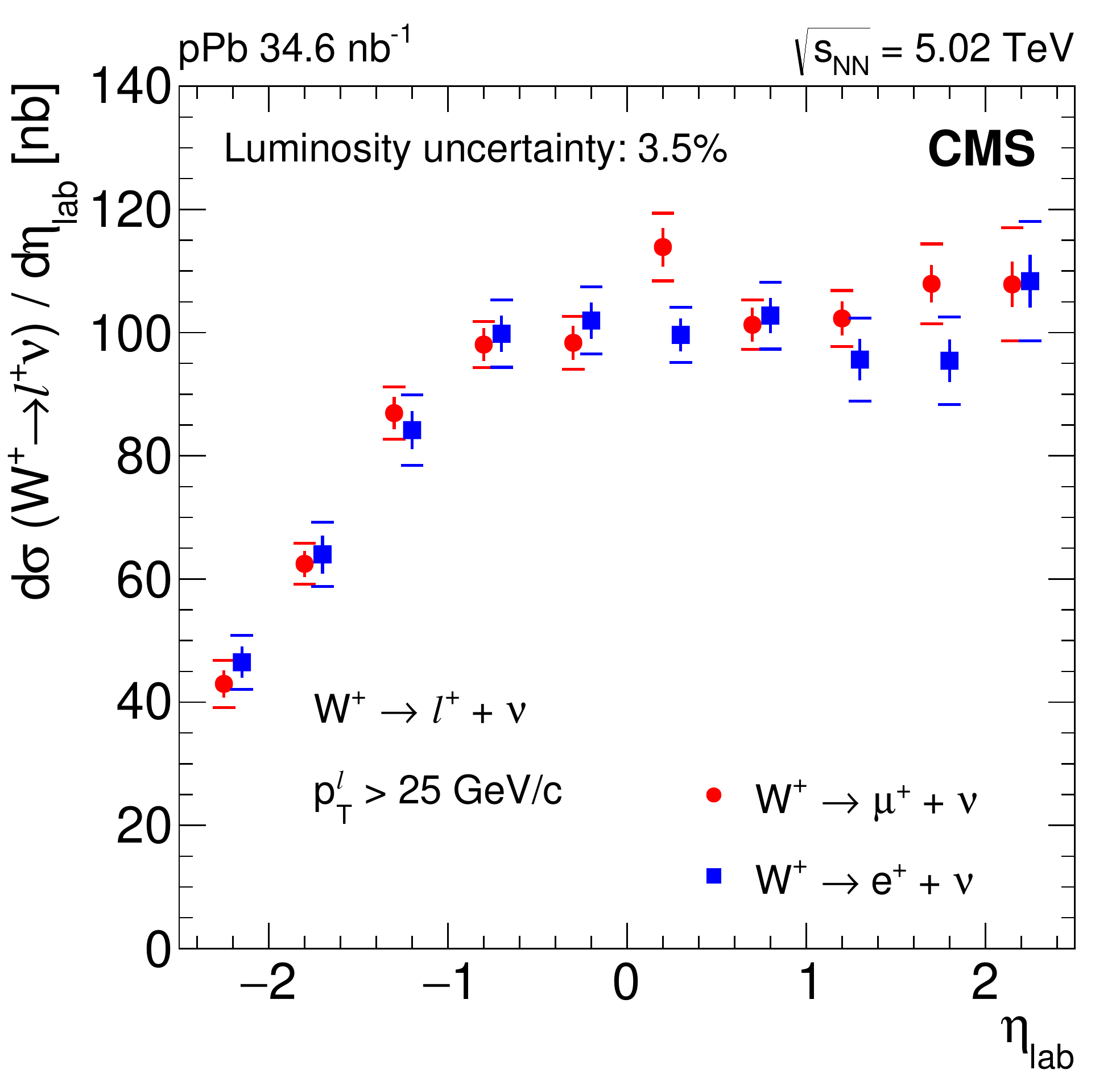}
 \includegraphics[width=0.45\textwidth]{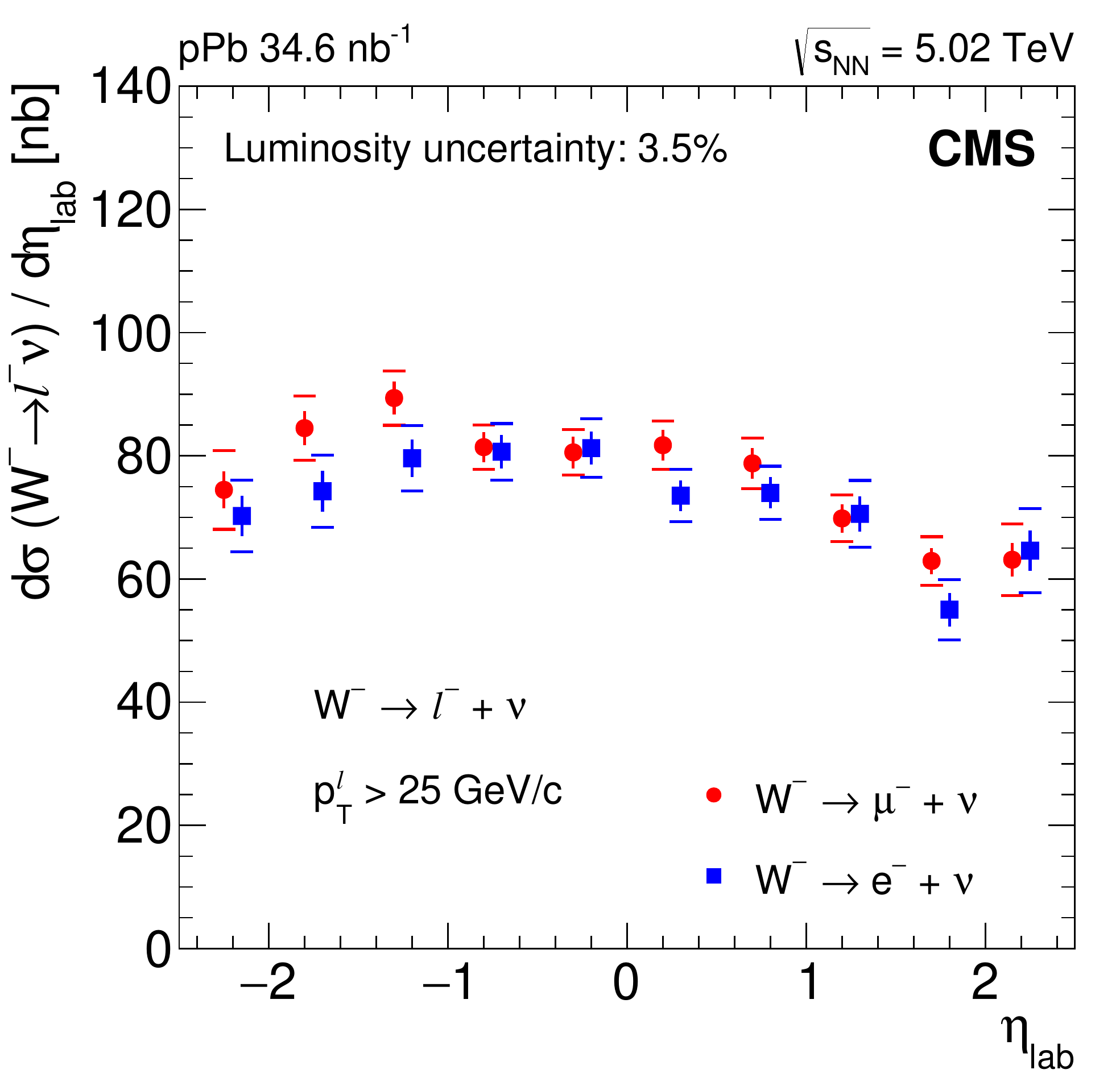}
 \caption{Production cross sections for $\PWp \to \ell^+ \nu$ (\cmsLeft) and $\PWm \to \ell^- \nu$ (\cmsRight),
as a function of the lepton pseudorapidity. Error bars represent the statistical uncertainties, while brackets show
    statistical and systematic uncertainties summed in quadrature. The global luminosity uncertainty of $\pm$3.5\% is not included. To improve visibility, the muon (electron) measurements, in red circles (blue squares), have been shifted by $-0.05$ ($+0.05$) in pseudorapidity.
}
  \label{FigYields}\end{figure}

Figure~\ref{FigYields} shows the production cross sections for $\pPb\to \PW^\pm + X \to \ell^\pm \nu + X$ as a function of the charged lepton pseudorapidity in the laboratory frame, with the lepton having $\pt > 25$\GeVc. The cross sections are determined by dividing the efficiency-corrected lepton yields by the integrated luminosity.

\begin{figure}[hbtp]\centering
 \includegraphics[width=0.45\textwidth]{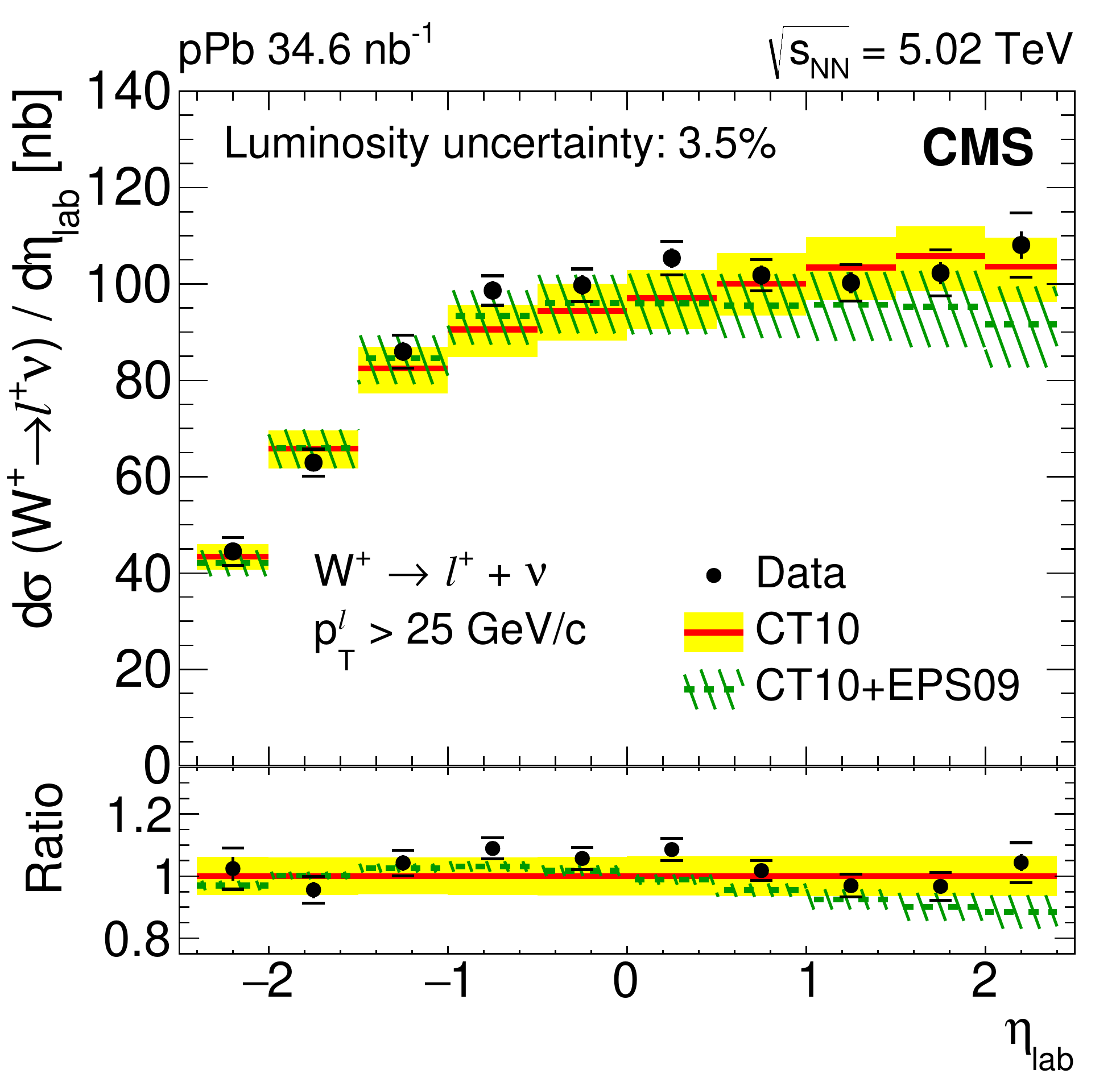}
 \includegraphics[width=0.45\textwidth]{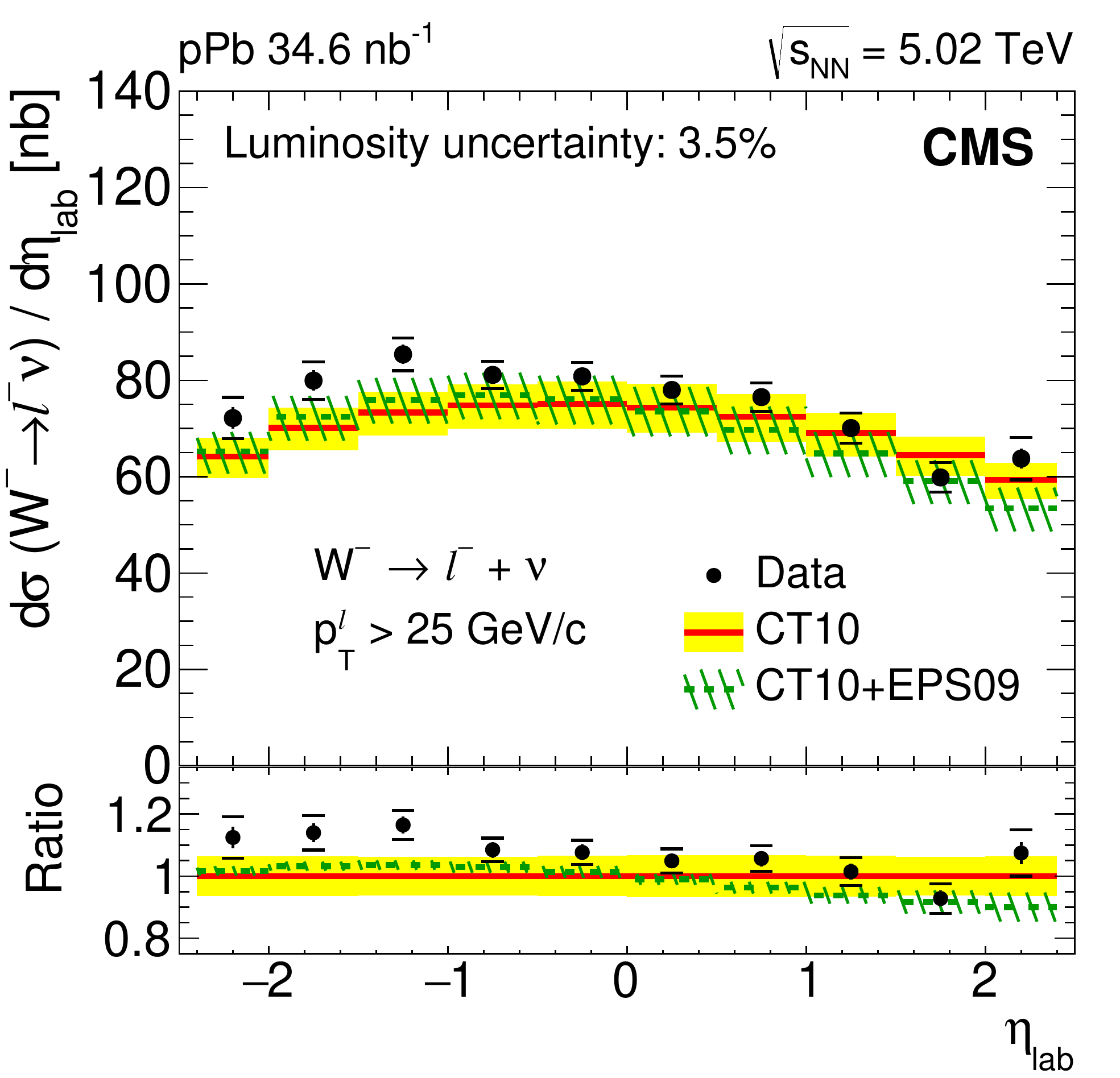}
 \caption{Production cross sections for $\PWp \to \ell^+ \nu$ (\cmsLeft) and $\PWm \to \ell^- \nu$ (\cmsRight),
as a function of the lepton pseudorapidity. Error bars represent the statistical uncertainties, while brackets show
    statistical and systematic uncertainties summed in quadrature. The global luminosity uncertainty of $\pm$3.5\% is not displayed.
Theoretical predictions with (CT10+EPS09, dashed green line) and without (CT10, solid red line)
PDF nuclear modifications are also shown, with the uncertainty bands.
The bottom panels show the ratio of the data (black points) and CT10+EPS09 (dashed green line) to the CT10 baseline. All theory uncertainty bands include scale and PDF uncertainties, except the EPS09 of the bottom panels which only includes the EPS09 PDF uncertainties.}
  \label{FigYieldsCombo}\end{figure}

Since the cross sections measured in the electron and muon channels
are found to be in good agreement with each other, they are combined using the BLUE method~\cite{blue_lyons}. Figure~\ref{FigYieldsCombo} compares the combined cross sections with next-to-leading-order (NLO) perturbative QCD predictions
provided by the authors of Ref.~\cite{Paukkunen:2010qg} using CT10~\cite{Lai:2010vv} proton parton distribution functions~(PDF) without or with EPS09~\cite{Eskola:2009uj} nPDF corrections, termed CT10 and CT10+EPS09, respectively. Their uncertainties are estimated as prescribed in Refs.~\cite{Lai:2010vv,Eskola:2009uj}.
Table~\ref{Cross-sectionMuPlus} gives the measured cross sections for each channel separately and combined, as a function of the lepton pseudorapidity, for positive and negative leptons. The theoretical predictions and their uncertainties (coming from the PDF set and from the renormalisation and factorisation scales) are also given.
The agreement between the data and both theoretical predictions is
within the uncertainties, although a small excess of
$\PW^-$ candidates appears at negative $\etalab$, \ie in the Pb ion beam direction.

The comparison between the CT10 and CT10+EPS09 calculations shows that the predicted modifications of the PDFs are of the same order as the theoretical uncertainties.
This indicates that cross sections alone lack discriminating power, and motivates the study of various asymmetries of the
$\ell^+$ and $\ell^-$ cross sections. The interest in such asymmetries is twofold. First, some of the experimental (e.g. integrated luminosity) and  theoretical (e.g. scale dependence) uncertainties cancel in such asymmetries. Second, the various asymmetries exhibit different sensitivities to the nuclear modifications of the PDFs, as discussed below.

\begin{table*}[htb]
\topcaption{\label{Cross-sectionMuPlus} Production cross section for $\pPb \to \PW + X \to \ell \nu + X$ for positively (top) and negatively (bottom) charged leptons of \pt larger than 25\GeVc, in nanobarns, as a function of the lepton pseudorapidity. Values are given first for muons and electrons separately, then combined. Quoted uncertainties are first statistical, then systematic.
Theoretical predictions with (CT10+EPS09) and without (CT10) PDF nuclear modifications are also given, with their uncertainties.
The global normalization uncertainty of 3.5\% is not included in the listed uncertainties.}
\centering
\resizebox{\textwidth}{!}{
\begin{tabular}{c|ccccc}
\hline
$\frac{\rd\sigma}{\rd\eta}$ (nb) [$\eta$ bin] & $[-2.4,-2.0]$ & $[-2.0,-1.5]$& $[-1.5,-1.0]$ & $[-1.0,-0.5]$ & $[-0.5,0]$\\
\hline
$\Pgmp$ & $43.0\pm2.2\pm3.1$ & $62.5\pm2.1\pm2.6$ & $86.9\pm2.6\pm3.4$ & $98.1\pm2.7\pm2.6$ & $98.3\pm2.8\pm3.3$\\
$\Pep$ & $46.5\pm2.6\pm3.6$ & $64.0\pm3.1\pm4.2$ & $84.2\pm3.1\pm4.8$ & $99.8\pm3.0\pm4.6$ & $102.0\pm2.9\pm4.6$\\
$\ell^+$ & $44.5\pm1.7\pm2.3$ & $62.9\pm1.8\pm2.2$ & $85.9\pm2.0\pm2.7$ & $98.6\pm2.1\pm2.3$ & $99.7\pm2.1\pm2.7$\\
\hline
CT10+EPS09 & $42.1^{+2.6}_{-2.8}$ & $66.0^{+3.8}_{-4.2}$ & $84.6^{+4.8}_{-5.4}$ & $93.4^{+5.3}_{-6.0}$ & $96.0^{+5.8}_{-6.3}$\\
CT10 & $43.4^{+2.5}_{-2.8}$ & $65.8^{+3.7}_{-4.2}$ & $82.4^{+4.6}_{-5.2}$ & $90.5^{+5.1}_{-5.7}$ & $94.4^{+5.7}_{-6.1}$\\
\hline \hline
$\frac{d\sigma}{d\eta}$ (nb) [$\eta$ bin] &$[0,0.5]$ & $[0.5,1.0]$ & $[1.0,1.5]$ & $[1.5,2.0]$& $[2.0,2.4]$ \\
\hline
$\Pgmp$ & $113.9\pm3.1\pm4.5$ & $101.3\pm2.8\pm2.9$ & $102.3\pm2.8\pm3.6$ & $107.9\pm3.1\pm5.7$ & $107.8\pm3.7\pm8.4$\\
$\Pep$ & $99.6\pm2.7\pm3.6$ & $102.8\pm2.9\pm4.6$ & $95.6\pm3.4\pm5.8$ & $95.4\pm3.5\pm6.2$ & $108.3\pm4.3\pm8.7$\\
$\ell^+$ & $105.3\pm2.1\pm2.8$ & $101.8\pm2.1\pm2.5$ & $100.2\pm2.2\pm3.1$ & $102.3\pm2.3\pm4.2$ & $108.1\pm2.8\pm6.0$\\
\hline
CT10+EPS09 & $95.9^{+6.2}_{-6.4}$ & $95.5^{+6.6}_{-6.7}$ & $95.7^{+6.8}_{-7.5}$ & $95.3^{+7.5}_{-8.4}$ & $91.6^{+7.9}_{-8.9}$\\
CT10 & $97.0^{+5.8}_{-6.4}$ & $100.0^{+6.4}_{-6.6}$ & $103.4^{+6.3}_{-6.8}$ & $105.7^{+6.2}_{-7.2}$ & $103.6^{+6.0}_{-7.3}$\\

\hline
\hline
$\frac{\rd\sigma}{\rd\eta}$\,(nb) [$\eta$ bin] & $[-2.4,-2.0]$ & $[-2.0,-1.5]$& $[-1.5,-1.0]$ & $[-1.0,-0.5]$ & $[-0.5,0]$\\
\hline
$\Pgmm$ & $74.5\pm3.0\pm5.6$ & $84.5\pm2.8\pm4.4$ & $89.4\pm2.6\pm3.5$ & $81.4\pm2.5\pm2.6$ & $80.6\pm2.6\pm2.6$\\
$\Pem$ & $70.2\pm3.2\pm4.8$ & $74.3\pm3.3\pm4.8$ & $79.6\pm3.1\pm4.3$ & $80.7\pm2.7\pm3.7$ & $81.3\pm2.6\pm4.0$\\
$\ell^-$ & $72.1\pm2.2\pm3.7$ & $79.9\pm2.1\pm3.3$ & $85.4\pm2.0\pm2.7$ & $81.1\pm1.8\pm2.1$ & $80.8\pm1.9\pm2.2$\\
\hline
CT10+EPS09 & $65.2^{+4.0}_{-4.6}$ & $72.4^{+4.4}_{-5.0}$ & $75.9^{+4.6}_{-4.9}$ & $76.9^{+4.6}_{-5.0}$ & $76.1^{+4.9}_{-5.3}$\\
CT10 & $64.2^{+3.9}_{-4.4}$ & $70.1^{+4.2}_{-4.7}$ & $73.3^{+4.3}_{-4.8}$ & $74.8^{+4.4}_{-4.8}$ & $75.1^{+4.7}_{-5.1}$\\
\hline \hline
$\frac{\rd\sigma}{\rd\eta}$\,(nb) [$\eta$ bin] &$[0,0.5]$ & $[0.5,1.0]$ & $[1.0,1.5]$ & $[1.5,2.0]$& $[2.0,2.4]$ \\
\hline
$\Pgmm$ & $81.7\pm2.5\pm3.0$ & $78.8\pm2.5\pm3.3$ & $69.8\pm2.3\pm3.0$ & $62.9\pm2.1\pm3.3$ & $63.1\pm2.8\pm5.1$\\
$\Pem$ & $73.5\pm2.5\pm3.5$ & $74.0\pm2.5\pm3.5$ & $70.6\pm2.8\pm4.6$ & $55.0\pm2.7\pm4.1$ & $64.6\pm3.3\pm6.0$\\
$\ell^-$ & $78.0\pm1.8\pm2.3$ & $76.5\pm1.8\pm2.4$ & $70.1\pm1.8\pm2.5$ & $59.8\pm1.7\pm2.6$ & $63.7\pm2.1\pm3.9$\\
\hline
CT10+EPS09 & $73.6^{+5.1}_{-5.2}$ & $69.7^{+4.9}_{-5.1}$ & $64.8^{+4.5}_{-4.9}$ & $59.1^{+4.3}_{-4.8}$ & $53.4^{+4.3}_{-4.8}$\\
CT10 & $74.3^{+4.9}_{-5.2}$ & $72.4^{+4.8}_{-5.1}$ & $69.1^{+4.2}_{-4.9}$ & $64.5^{+3.8}_{-4.3}$ & $59.3^{+3.6}_{-4.0}$\\
\hline
\end{tabular}
}
\end{table*}

\begin{figure}[hbtp]\centering
 \includegraphics[width=\cmsFigWidth]{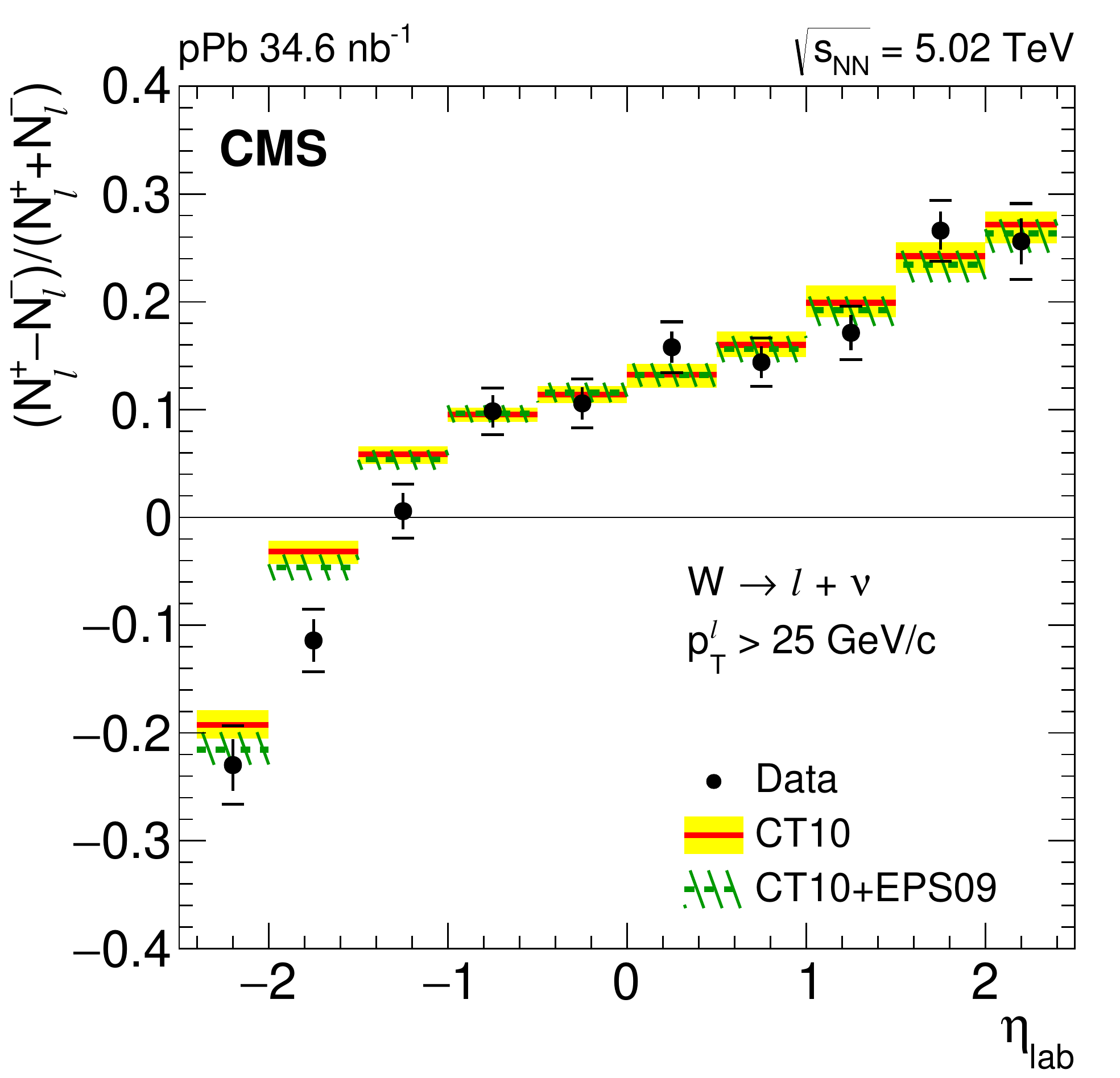}
 \caption{Lepton charge asymmetry, $(N_{\ell}^+ - N_{\ell}^-)/(N_{\ell}^+ + N_{\ell}^-)$, as a function of the lepton pseudorapidity. Error bars represent the statistical uncertainties, while brackets show
    statistical and systematic uncertainties summed in quadrature. Theoretical predictions with (CT10+EPS09, dashed green line) and without (CT10, solid red line) PDF nuclear modifications are also shown, with their uncertainty bands.}
  \label{FigChargeAsymmetry}\end{figure}

The lepton charge asymmetry, defined as $(N_{\ell}^+ - N_{\ell}^-)/(N_{\ell}^+ + N_{\ell}^-)$ with $N_{\ell}^\pm$ being the efficiency-corrected lepton yields, is shown in
Fig.~\ref{FigChargeAsymmetry}, as a function of $\etalab$, and compared to the theoretical predictions.
For $\etalab > -1$, both calculations reproduce the present measurements. For $\etalab < -1$, however, the two calculations overpredict the asymmetry values.
A possible physical origin of this disagreement could be a different modification of \cPqu\ and \cPqd\ quark distributions in nuclei. In proton-(anti)proton collisions, the W-boson charge asymmetry  is known to be a sensitive probe of the down-to-up quark PDF ratio in a proton, $\cPqd^p/\cPqu^p$~\cite{Aaltonen:2009ta,Chatrchyan:2013mza,Chatrchyan:2012xt}.
Similarly, this asymmetry in pPb collisions measured in the lead fragmentation region (\ie $\etalab<0.465$) probes these quark densities in a nucleon inside the lead nucleus. Assuming the standard isospin symmetry ($\cPqu^p = \cPqd^n$, $\cPqu^n = \cPqd^p$), one can define a similar ratio, $\cPqd^{p/A}/\cPqu^{p/A}=\cPqd^{p}/\cPqu^{p} \times R_\cPqd/R_\cPqu$, where $R_i$ are the nPDF ratios, $R_\cPqu \equiv \cPqu^{p/A}/\cPqu^p$ and $R_\cPqd \equiv \cPqd^{p/A}/\cPqd^p$. The typical quark momentum fraction probed in the Pb nucleus is given by $x\simeq M_W/\rootsNN\times \exp(-\etalab+0.465)$ (assuming that the \PW\ boson rapidity is similar to that of the lepton), therefore $x\simeq 0.02-0.20$ in the range $-2 < \etalab < 0$. In most global fit analyses of the nPDFs (as in the case of EPS09), it is assumed that the nuclear ratios respect the isospin symmetry, namely $R_\cPqu=R_\cPqd$, essentially to minimise the number of free parameters in the fits. However, no physical reason prevents nuclear modifications to be different for up and down quark PDFs. For example, it is known that the shapes of the up and down quark distributions in protons are different~\cite{Martin:2009iq}.
Furthermore, the present disparity between data and theory is unlikely to come from the proton PDF assumption, given the excellent agreement of lepton charge asymmetry measured in pp collisions by CMS~\cite{Chatrchyan:2013mza} and ATLAS~\cite{Aad:2011yna} with NLO calculations using CT10 parton densities.

\begin{figure}[hbtp]\centering
 \includegraphics[width=0.45\textwidth]{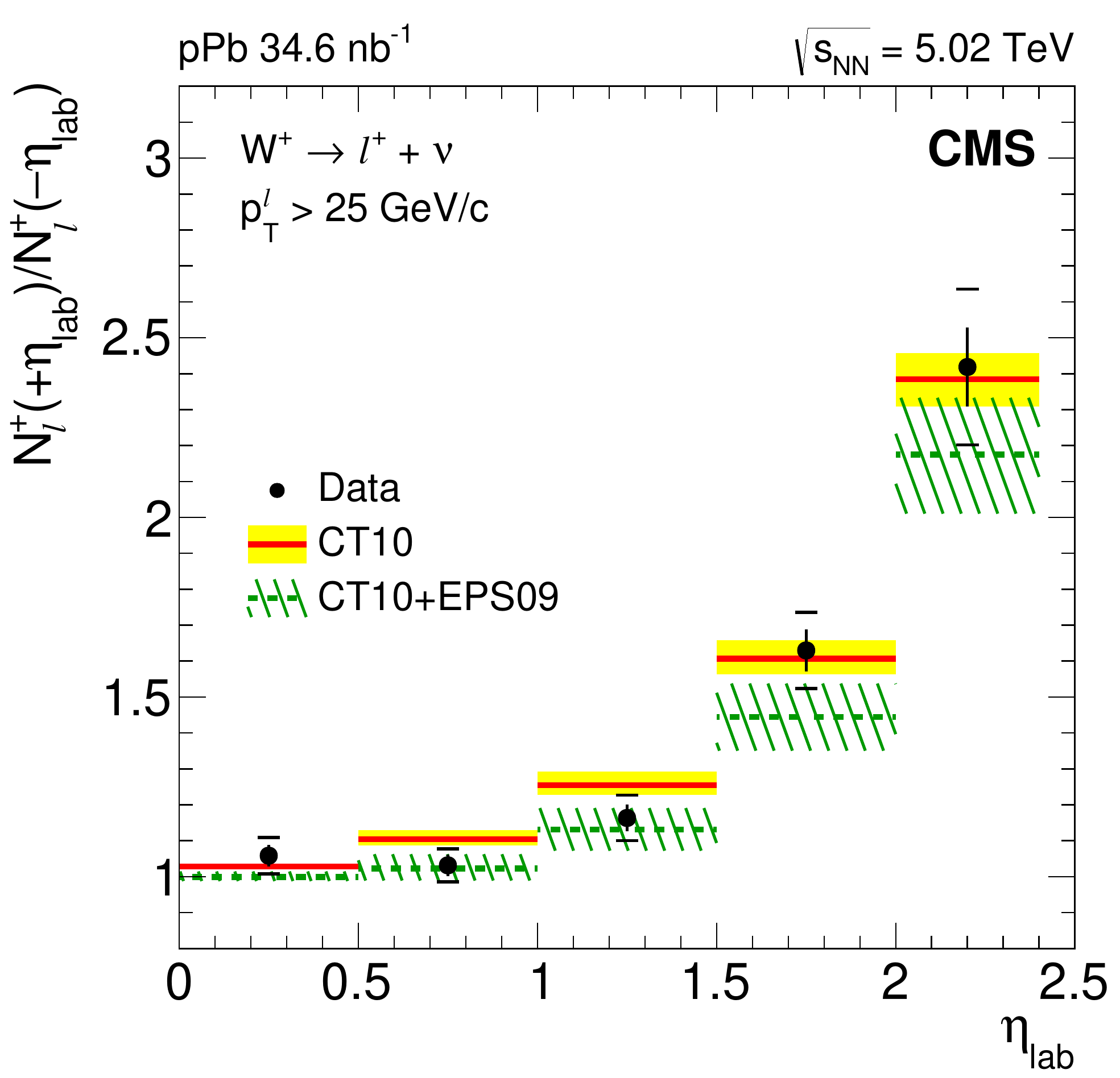}
 \includegraphics[width=0.45\textwidth]{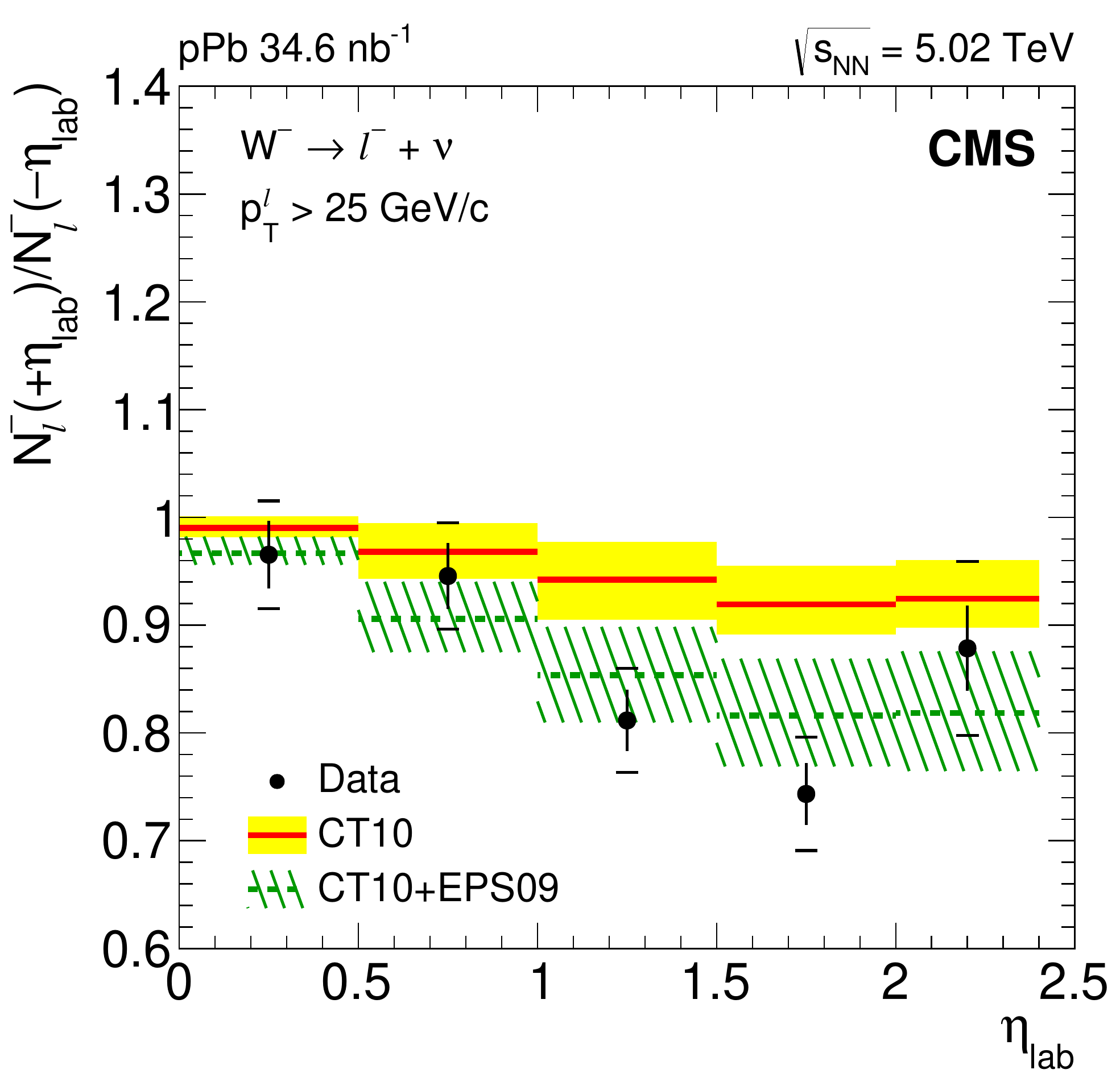}
 \caption{Forward-backward asymmetries, $N_{\ell}(+\etalab)/N_{\ell}(-\etalab)$, for the positive (\cmsLeft) and negative (\cmsRight)
leptons. Error bars represent the statistical uncertainties, while brackets show
    statistical and systematic uncertainties summed in quadrature. Theoretical predictions with (CT10+EPS09, dashed green line) and without (CT10, solid red line) PDF nuclear modifications are also shown, with their uncertainty bands.}
  \label{FigA1}\end{figure}

A traditional way to probe nuclear parton densities is to compare the forward and backward W yields, that are respectively sensitive to the nPDFs at small and large $x$. The forward-backward asymmetries $N_{\ell}^{\pm}(+\etalab)/N_{\ell}^{\pm}(-\etalab)$ are shown in Fig.~\ref{FigA1}, separately
for the positively and negatively charged leptons, and compared to the same predictions as mentioned above.
Given the experimental accuracy and the magnitude of the differences between the two sets of predictions, the measurements have a potential to discriminate between them.
However, although the negative lepton decay channel appears to slightly favour the CT10+EPS09 prediction over the CT10 calculation, the positive lepton channel does not, thus no firm conclusion can be drawn.

\begin{figure}[hbt]\centering
 \includegraphics[width=\cmsFigWidth]{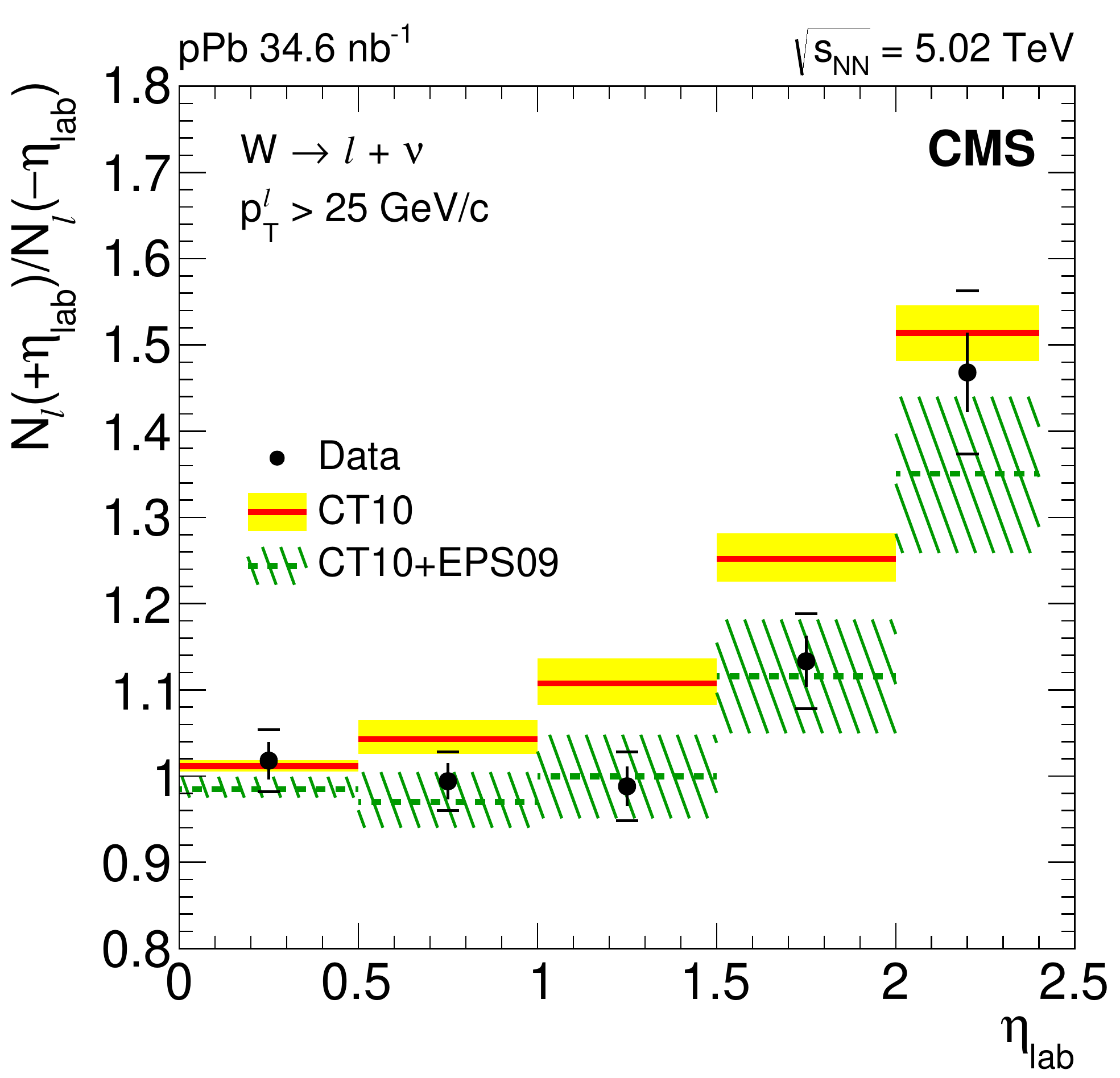}
 \caption{The forward-backward asymmetry of charge-summed \PW\ bosons, as a function of the lepton pseudorapidity. Error bars represent the statistical uncertainties, while brackets show
    statistical and systematic uncertainties summed in quadrature. Theoretical predictions with (CT10+EPS09, dashed green line) and without (CT10, solid red line) PDF nuclear modifications are also shown, with their uncertainty bands.}
  \label{FigA3}\end{figure}

Another asymmetry variable, $(N_{\ell}^+(+\etalab) - N_{\ell}^+(-\etalab)) / (N_{\ell}^-(+\etalab) - N_{\ell}^-(-\etalab))$, was proposed in
Ref.~\cite{Paukkunen:2010qg} to reach maximum sensitivity to nuclear modifications of PDFs. However, this asymmetry probability distribution shows
a very non-Gaussian behaviour, when its denominator approaches zero, and its sign can be flipped within the
uncertainty. A different asymmetry is proposed here, $N_{\ell}(+\etalab)/ N_{\ell}(-\etalab)$, a forward-backward asymmetry
of the charge-summed W bosons, which achieves a similar sensitivity.
As in the case of the charge asymmetry, this asymmetry can be related to the nuclear modifications of the PDFs within the lead nucleus. Here, forward (backward) W boson production is sensitive to the PDFs of the sea quark at $x\sim 10^{-3}$ (valence quark at $x\sim 10^{-1}$) in the lead nucleus. Therefore, the forward-backward ratio probes the small-$x$ modification of the lead nucleus PDF (shadowing) over the large-$x$ modifications (anti-shadowing).
This asymmetry is shown in Fig.~\ref{FigA3}, and deviates from unmodified PDFs, more clearly favouring CT10+EPS09 over CT10.

In order to quantify the agreement between the data and the expectation from the CT10 and CT10+EPS09 calculations, a $\chi^2$ test is performed
for each of the above (correlated) variables.
The few correlations in experimental uncertainties described above, only relevant for $\PW^\pm$ boson cross sections but not for asymmetries,
are taken into account, as well as the correlations in theoretical uncertainties. The resulting $\chi^2$ values and probabilities
are given in Table~\ref{tab:chi2}. The CT10+EPS09 calculations provide a better description of the data, with still a relatively low probability for the lepton charge asymmetry, because of the backward region.

\begin{table*}
 \topcaption{Values of the $\chi^2$ test between the measurements and the theoretical predictions, with (CT10+EPS09) or without (CT10) nuclear
    modifications of the PDFs. The probability (Prob.) to measure a value greater to that measured in data is also given for ten degrees of freedom in the case of the first three observables and five degrees of freedom for the three others observables.
\label{tab:chi2}}
 \centering
\renewcommand{\arraystretch}{1.3}
\begin{tabular}{ccc@{\hspace{1em}}c@{\hspace{1em}}cc}
\hline
\multicolumn{1}{c}{\multirow{2}{*}{Observable}}&\multicolumn{2}{c}{CT10}&& \multicolumn{2}{c}{CT10+EPS09}\\ \cline{2-3} \cline{5-6}
& $\chi^2$ & Prob.(\%) && $\chi^2$ & Prob.(\%) \\ \cline{2-3}
\hline
$\rd\sigma / \rd\eta (\ell^+)$ & 13 & 25 && 8.6 & 57 \\
$\rd\sigma / \rd\eta (\ell^-)$ & 15 & 14 && 8.2 & 60 \\
$(N_{\ell}^+ - N_{\ell}^-) / (N_{\ell}^+ + N_{\ell}^-)$ & 15 & 12 && 11 & 35 \\
$N_{\ell}^+ (+\eta_\text{lab}) / N_{\ell}^+ (-\eta_\text{lab})$ & 3.1 & 68 && 3.2 & 68 \\
$N_{\ell}^- (+\eta_\text{lab}) / N_{\ell}^- (-\eta_\text{lab})$ & 9.7 & 8.4 && 3.5 & 63 \\
$N_{\ell} (+\eta_\text{lab}) / N_{\ell} (-\eta_\text{lab})$ & 6.2 & 29 && 2.1 & 83 \\
\hline
\end{tabular}
\end{table*}

\section{Summary}

The first measurement of W boson production in pPb collisions has been reported, using the electron and muon decay modes for leptons of \pt above 25\GeVc and $\abs{\etalab}< 2.4$. The differential cross sections as a function
of the lepton pseudorapidity agree with theoretical predictions assuming both unmodified (CT10) and modified (CT10+EPS09) nPDFs, except in the most backward region (Pb ion beam direction), where a hint of an enhancement is seen for the \PWm\ bosons.
In the same region, the related lepton charge asymmetry deviates slightly from the predictions, something that could potentially arise from different nuclear modifications of the up and down quark PDFs.
In a related observation, forward-backward asymmetries show a deviation from unmodified PDFs. Taken together, these measurements
show the need for including W boson data in
 nuclear parton distribution global fits.

\begin{acknowledgments}

We congratulate our colleagues in the CERN accelerator departments for the excellent performance of the LHC and thank the technical and administrative staffs at CERN and at other CMS institutes for their contributions to the success of the CMS effort. In addition, we gratefully acknowledge the computing centres and personnel of the Worldwide LHC Computing Grid for delivering so effectively the computing infrastructure essential to our analyses. Finally, we acknowledge the enduring support for the construction and operation of the LHC and the CMS detector provided by the following funding agencies: BMWFW and FWF (Austria); FNRS and FWO (Belgium); CNPq, CAPES, FAPERJ, and FAPESP (Brazil); MES (Bulgaria); CERN; CAS, MoST, and NSFC (China); COLCIENCIAS (Colombia); MSES and CSF (Croatia); RPF (Cyprus); MoER, ERC IUT and ERDF (Estonia); Academy of Finland, MEC, and HIP (Finland); CEA and CNRS/IN2P3 (France); BMBF, DFG, and HGF (Germany); GSRT (Greece); OTKA and NIH (Hungary); DAE and DST (India); IPM (Iran); SFI (Ireland); INFN (Italy); MSIP and NRF (Republic of Korea); LAS (Lithuania); MOE and UM (Malaysia); CINVESTAV, CONACYT, SEP, and UASLP-FAI (Mexico); MBIE (New Zealand); PAEC (Pakistan); MSHE and NSC (Poland); FCT (Portugal); JINR (Dubna); MON, RosAtom, RAS and RFBR (Russia); MESTD (Serbia); SEIDI and CPAN (Spain); Swiss Funding Agencies (Switzerland); MST (Taipei); ThEPCenter, IPST, STAR and NSTDA (Thailand); TUBITAK and TAEK (Turkey); NASU and SFFR (Ukraine); STFC (United Kingdom); DOE and NSF (USA).

Individuals have received support from the Marie-Curie programme and the European Research Council and EPLANET (European Union); the Leventis Foundation; the A. P. Sloan Foundation; the Alexander von Humboldt Foundation; the Belgian Federal Science Policy Office; the Fonds pour la Formation \`a la Recherche dans l'Industrie et dans l'Agriculture (FRIA-Belgium); the Agentschap voor Innovatie door Wetenschap en Technologie (IWT-Belgium); the Ministry of Education, Youth and Sports (MEYS) of the Czech Republic; the Council of Science and Industrial Research, India; the HOMING PLUS programme of the Foundation for Polish Science, cofinanced from European Union, Regional Development Fund; the Compagnia di San Paolo (Torino); the Consorzio per la Fisica (Trieste); MIUR project 20108T4XTM (Italy); the Thalis and Aristeia programmes cofinanced by EU-ESF and the Greek NSRF; and the National Priorities Research Program by Qatar National Research Fund.
\end{acknowledgments}
\bibliography{auto_generated}

\cleardoublepage \appendix\section{The CMS Collaboration \label{app:collab}}\begin{sloppypar}\hyphenpenalty=5000\widowpenalty=500\clubpenalty=5000\textbf{Yerevan Physics Institute,  Yerevan,  Armenia}\\*[0pt]
V.~Khachatryan, A.M.~Sirunyan, A.~Tumasyan
\vskip\cmsinstskip
\textbf{Institut f\"{u}r Hochenergiephysik der OeAW,  Wien,  Austria}\\*[0pt]
W.~Adam, T.~Bergauer, M.~Dragicevic, J.~Er\"{o}, M.~Friedl, R.~Fr\"{u}hwirth\cmsAuthorMark{1}, V.M.~Ghete, C.~Hartl, N.~H\"{o}rmann, J.~Hrubec, M.~Jeitler\cmsAuthorMark{1}, W.~Kiesenhofer, V.~Kn\"{u}nz, M.~Krammer\cmsAuthorMark{1}, I.~Kr\"{a}tschmer, D.~Liko, I.~Mikulec, D.~Rabady\cmsAuthorMark{2}, B.~Rahbaran, H.~Rohringer, R.~Sch\"{o}fbeck, J.~Strauss, W.~Treberer-Treberspurg, W.~Waltenberger, C.-E.~Wulz\cmsAuthorMark{1}
% ...
\vskip\cmsinstskip
% ...
1:~~Also at Vienna University of Technology, Vienna, Austria\\
2:~~Also at CERN, European Organization for Nuclear Research, Geneva, Switzerland\\
% ...

\end{sloppypar}
\end{document}